\documentstyle[12pt]{article}
\newcounter{appendixc}
\newcounter{subappendixc}[appendixc]
\newcounter{subsubappendixc}[subappendixc]

\renewcommand{\appendix}[1] {\vspace*{0.6cm}
        \refstepcounter{appendixc}
        \setcounter{figure}{0}
        \setcounter{table}{0}
        \setcounter{equation}{0}
        \renewcommand{\thefigure}{\Alph{appendixc}.\arabic{figure}}
        \renewcommand{\thetable}{\Alph{appendixc}.\arabic{table}}
        \renewcommand{\theappendixc}{\Alph{appendixc}}
        \renewcommand{\theequation}{\Alph{appendixc}.\arabic{equation}}
        \noindent{\bf Appendix \theappendixc #1}\par\vspace*{0.4cm}}

\begin{document}
\begin{titlepage}
\begin{flushright}
NUHEP-TH-96-6\\
\end{flushright}
\vspace{0.1in}
\baselineskip=0.25in
\begin{center}
{\Large Supersymmetric Electroweak Corrections to Single \\ 
              Top Quark Production at the Fermilab Tevatron }

\vspace{.2in}

     Chong Sheng Li{\footnote {On leave from Department of Physics,
                           Peking University, China

                ~~e-mail: csli@nuhep.phys.nwu.edu

                ~~(After December: csli@svr.bimp.pku.edu.cn)}},
     Robert J. Oakes {\footnote{ e-mail: oakes@fnal.gov}}
and  Jin Min Yang{\footnote  {On leave from Department of Physics,
                              Henan Normal University, China

                              ~~e-mail: jmyang@nuhep.phys.nwu.edu}}

\vspace{.2in}

     Department of Physics and Astronomy, Northwestern University,\\
     Evanston, Illinois 60208-3112, USA

\end{center}
\vspace{.4in}

\begin{footnotesize}
\begin{center}\begin{minipage}{5in}
\baselineskip=0.25in

\begin{center} ABSTRACT\end{center}
                    
  We have calculated the $O(\alpha_{ew} M_t^2/M_W^2)$ 
supersymmetric electroweak corrections to single top quark 
production via $q \bar q'\rightarrow t \bar b$
at the Fermilab Tevatron in the minimal supersymmetric model.
The supersymmetric electroweak corrections to the cross section
are a few percent for $\tan \beta>1$, and can exceed 
10\% for $\tan\beta<1$.
The combined effects of SUSY electroweak 
corrections and  the Yukawa corrections
can exceed 10\%  for favorable parameter 
values, which might be observable at a high-luminosity Tevatron. 
\end{minipage}\end{center}
\end{footnotesize}
\vspace{.5in}

PACS number: 14.80Dq; 12.38Bx; 14.80.Gt

\end{titlepage}
\eject
\baselineskip=0.25in
\begin{center} {\Large 1. Introduction }\end{center}

The top quark has now been discovered by the CDF and D0 
collaborations at the Fermilab Tevatron[1]. 
Measurements of its mass are $176\pm 9 GeV$ and 
$170\pm 18$ from CDF and D0, respectively, and the 
world-average value of the top
quark mass from Run I at the Tevatron was recently  reported to be $175\pm 6 
GeV$, based on roughly 100 $pb^{-1}$ of data[2]. At the Tevatron the dominant 
production mechanism of the top quark is the QCD pair production process
$q \bar q \rightarrow t \bar t$[3]. However, because the top quark is so heavy, 
electroweak production of single top quarks can become significant, 
particularly at the next Tevatron
Run-II.  With $\sqrt s = 2$ TeV and an integrated luminosity of $2 fb^{-1}$ 
one can expect, in the Standard Model(SM), that for a $175GeV$ top quark, 
there will be about $1.4 \times 10^4 t\bar t$ pairs and $5 \times 10^3$ 
single top events produced[4], which is about $35\%$ of the total $t\bar t$ 
rate. After taking into account the $b$-tagging efficiency and the detection 
efficiency[5], there are about 1000 single-$b$-tagged $t\bar t$ pairs in the 
$l$+jets sample, 100 in the dilepton sample, and 250 single top events in the 
$l$+jets sample available for testing various properties of the top quark.
Even with fewer events, single top production processes are 
important because they involve the electroweak interaction and, therefore,
can probe the electroweak sector of the theory, in contrast to the QCD pair
production mechanism, and provide a consistency check on the measured
parameters of the top quark in the QCD pair production. At the Tevatron
single top quarks are produced primarily via the $W$-gluon fusion process[6]
and the Drell-Yan type single top process, 
$q \bar q'\rightarrow W^* \rightarrow t\bar b$($W^*$ process)[7], 
which can reliably be 
predicted in the SM. The theoretical uncertainty in the cross section 
is only about a few percent due to QCD corections[8].
As analysed in Ref.[9],
 the statistical error in the measured cross section for the $W^*$ process 
at the Tevatron will be about $\pm 30\%$; however, a high-luminosity Tevatron 
would allow a measurement of the cross section with a statistical uncertainty 
of about 6\%[9]. At this level of experimental accuracy a calculation of the 
radiative corrections in the SM is necessary  and effects beyond the SM,
for example SUSY corrections, should also be considered. 

In Ref.[9] the QCD and Yukawa corrections to the $W^*$ process
have been calculated in the SM. In a previous paper [10] we calculated the 
Yukawa corrections to this process from the Higgs sector in the general 
Two-Higgs-Doublet Model(2HDM) and the Minimal Supersymmetric 
Model(MSSM)[11] and found that the corrections can amount to more than a 15\%
reduction in the production cross section relative to the tree leve result
in the 2HDM, and a 10\% enhancement in the MSSM. However, in the MSSM, 
in addition to these Yukawa corrections from the 
Higgs sector, the Supersymmetric (SUSY) corrections due to super particles
(sparticles) should also be taken into account. The dominant virtual effects
of sparticles arise from the SUSY electroweak corrections of order 
$O(\alpha_{ew} M_t^2/M_W^2)$ and the SUSY QCD corrections of order
$O(\alpha_s)$ which arise from loops of charginos, neutralinos and squarks,
and gluinos and squarks. It is well-known[12] that the anomalous
magnetic moment for a spin $1/2$ fermion vanishes in the SUSY limit 
and away from the SUSY limit there is a partial cancellation.
Therefore, in general, one can expect the Yukawa corrections from the Higgs
sector and the SUSY electroweak corrections from virtual charginos
and neutralinos to cancel to some extent.  

In this paper we present the calculation of the $O(\alpha_{ew} M_t^2/M_W^2)$ 
SUSY electroweak corrections at the Fermilab Tevatron in the MSSM, 
and show the combined effect of including all the 
$O(\alpha_{ew} M_t^2/M_W^2)$ terms from both Yukawa corrections and
SUSY electroweak corrections.
The $O(\alpha_s)$ SUSY QCD corrections will be given elsewhere[13]. 
The paper is organized as follows: In Sec. II we present the analytic results
in terms of the well-known standard notation of one-loop Feynman integrals.
In Sec. III  we give some numerical examples and discuss the implications
of our reults.
\vspace{1cm}

\begin{center} {\Large 2. Calculations }\end{center}
\vspace{.3cm}

\begin{flushleft}{\large 2.1 Formalism}\end{flushleft}

 In our calculations we used dimensional regularization to 
control
all the ultraviolet divergences in the virtual loop corrections and we adopted
the on-mass-shell renormalization scheme[14].
Including the $O(\alpha_{ew} M_t^2/M_W^2)$ 
electroweak corrections,
the renormalized amplitude for
$q\bar q'\rightarrow t\bar b$ can be written as
\begin{equation}
M_{ren}=M_0 +\delta M^{SUSY-EW}
\end{equation}
where $M_0$ is  the tree-level matrix element
and $\delta M^{SUSY-EW}$ represents the  SUSY electroweak corrections. 
  The tree-level Feynman diagram for single top quark production via
$q \bar q'\rightarrow t \bar b$ is shown in Fig.1(a). The amplitude 
$M_0$ is given by
\begin{equation}
 M_0= i \frac{g^2}{2} \frac{1}{\hat s-M_W^2}
  \bar v(p_2) \gamma_{\mu} P_L u(p_1)  \bar u(p_3) \gamma^{\mu} P_L v(p_4).
\end{equation} 
The amplitude $\delta M^{SUSY-EW}$ is
given in the following section.
Here $p_1$ and $p_2$ denote the momentum of the incoming quarks $q$ and $\bar q'$,
while $p_3$ and $p_4$ are used for the outgoing $t$ and
$\bar b$ quarks, and $\hat{s}$ is the center-of-mass energy of the
subprocess.

  The renormalized differential cross section for the subprocess is 
\begin{equation}
\frac{d\hat{\sigma}}{d\cos\theta}=\frac{\hat{s}-M_t^2}{32\pi\hat s^2}
\overline{\sum} \vert M_{ren}\vert^2,
\end{equation}
where $\theta$ is the angle between the top quark and incoming quark.
Integrating this differential cross section over $\cos\theta$ 
one obtains the cross section for subprocess
\begin{equation}
\hat{\sigma}=\hat{\sigma}_0
	+\Delta \hat{\sigma}
\end{equation}
where the tree-level cross section $\hat{\sigma}_0$ is given by
\begin{equation}
\hat{\sigma}_0=\frac{g^4}{128\pi}\frac{\hat{s}-M^2_t}{\hat{s}^2(\hat{s}-
M^2_W)^2}[\frac{2}{3}(\hat{s}-M_t^2)^2
 +(\hat{s}-M^2_t)(M^2_t+M^2_b) + 2M^2_tM^2_b],
\end{equation}
and $\Delta \hat{\sigma}$ represents the SUSY electroweak corrections.

The total hadronic cross section for the single production of top quarks via 
$q\bar q'$ can be written in the form
\begin{equation}
\sigma (s)=\sum_{i,j}\int dx_1 dx_2 \hat\sigma_{ij}(x_1x_2s, M_t^2,
\mu^2)[f^A_i(x_1,\mu)f^B_j(x_2, \mu)+(A\leftrightarrow B)],
\end{equation}
where
\begin{eqnarray}
s&=&(P_1+P_2)^2,\\
\hat{s}&=&x_1x_2s,\\
p_1&=&x_1P_1,
\end{eqnarray}
and 
\begin{equation}
p_2=x_2P_2.~~~~~~~~
\end{equation}
Here $A$ and $B$ denote the incident hadrons and $P_1$ and $P_2$ are their
four-momenta, while $i, j$ are the initial partons and $x_1$ and $x_2$ are 
their longitudinal momentum fractions. The functions $f^A_i$ and $f^B_j$ are 
the usual parton distributions[15,16]. 
Finally, introducing the convenient variable 
$\tau =x_1x_2$ and changing independent variables, the total cross section
becomes
\begin{equation}
\sigma(s)=\sum_{i,j}\int^1_{\tau_0}\frac{d\tau}{\tau}(\frac{1}{s}
\frac{dL_{ij}}{d\tau})(\hat s \hat \sigma_{ij})
\end{equation}
where $\tau_0=(M_t+M_b)^2/s$. The quantity $dL_{ij}/d\tau$ is the parton
luminosity, which is defined to be
\begin{equation}
\frac{dL_{ij}}{d\tau}=\int^1_{\tau} \frac{dx_1}{x_1}[f^A_i(x_1,\mu)
f^B_j(\tau/x_1,\mu)+(A\leftrightarrow B)]
\end{equation}
\vspace{.4cm}

\begin{flushleft}{\large 2.2 SUSY electroweak corrections }\end{flushleft}

  The SUSY electroweak corrections of order $\alpha_{ew} M_t^2/M_W^2$ to
the process $q \bar q'\rightarrow t \bar b$ arise from the Feynman diagrams
shown in Fig.1(b)-(g). The matrix element for these corrections 
can be written as
\begin{equation}
\delta M^{SUSY-EW}=\delta M^{SUSY-EW}_{Wt\bar b}+\delta M_{box}
                  +\delta M_{box}^c
\end{equation}
where $\delta M^{SUSY-EW}_{Wt\bar b}$ represents corrections arising from
the self-energy diagrams and vertex diagrams [Fig.1(b)-(e)], 
while $\delta M_{box}$ and
$\delta M_{box}^c$ correspond to the box diagram [Fig1.(f)] and 
crossed box diagram[Fig1.(g)],
respectively. $\delta M^{SUSY-EW}_{Wt\bar b}$ is given by
\begin{eqnarray}
\delta M^{SUSY-EW}_{Wt\bar b}&= i& \frac{g^2}{2} \frac{1}{\hat s-M_W^2}
  \bar v(P_2) \gamma_{\mu} P_L u(P_1)
  \bar u(P_3) \left [\gamma^{\mu} P_L (\frac{1}{2}\delta Z^t_L
                                 +\frac{1}{2}\delta Z^b_L+E^L_1)\right.
        \nonumber\\
& &  \left.+P^{\mu}_3 P_L E^L_2+ P^{\mu}_4 P_L E^L_3\right ] v(P_4)
\end{eqnarray}
The renormalization constants and form factors in Eq.(14) are 
\begin{eqnarray}
\delta Z^t_L&=&
  \frac{1}{8\pi^2}\left [ \vert R_{\tilde t_i \tilde \chi^0_j}\vert^2
                    (-\frac{\Delta}{2}+F_1^{(t\tilde \chi^0_j \tilde t_i)})
         +2 M_t M_{\tilde \chi^0_j}L_{\tilde t_i \tilde \chi^0_j}
         R_{\tilde t_i \tilde \chi^0_j}^* G_0^{(t\tilde \chi^0_j \tilde t_i)}
        \right. \nonumber\\
& &\left.
+M_t^2(\vert R_{\tilde t_i \tilde \chi^0_j}\vert^2
               +\vert L_{\tilde t_i \tilde \chi^0_j}\vert^2)
                 G_1^{(t\tilde \chi^0_j \tilde t_i)}\right ]
 +\frac{g^2}{32\pi^2}\lambda_t^2 L_{\tilde b_i}^2\vert V_{j2}\vert^2
        M_t^2 G_1^{(t\tilde \chi^+_j \tilde b_i)},\\
\delta Z^b_L&=&\frac{g^2}{32\pi^2}\lambda_t^2 R_{\tilde t_i}^2
  \vert V_{j2}\vert^2(-\frac{\Delta}{2}+F_1^{(b\tilde \chi^+_j \tilde t_i)}),\\
E^L_1&=&\frac{g}{8\sqrt 2\pi^2}\lambda_t V_{j2}^* R_{\tilde t_i}\left\{
 R_{\tilde t_i \tilde \chi^0_k}O^{R^*}_{kj}M_{\tilde \chi^+_j}M_{\tilde 
\chi^0_k}c_0
+L_{\tilde t_i \tilde \chi^0_k}O^{R^*}_{kj}M_{\tilde \chi^+_j}M_t(c_0+c_{12})
        \right. \nonumber \\
& &\left.+L_{\tilde t_i \tilde \chi^0_k}O^{L^*}_{kj}M_{\tilde \chi^0_k}M_t 
c_{12}+R_{\tilde t_i \tilde \chi^0_k}O^{L^*}_{kj}[M_t^2(c_{22}-c_{23})
+\hat s(c_{12}+c_{23})+2c_{24}-\frac{1}{2}]\right\},\\
E^L_2&=&-\frac{g}{4\sqrt 2\pi^2}\lambda_t V_{j2}^* R_{\tilde t_i}
        O^{L^*}_{kj}\left[
 L_{\tilde t_i \tilde \chi^0_k}M_{\tilde \chi^0_k} c_{12}
+R_{\tilde t_i \tilde \chi^0_k}M_t(c_{12}+c_{22})\right ],
\end{eqnarray}
and
\begin{equation}
E^L_3=\frac{g}{4\sqrt 2\pi^2}\lambda_t V_{j2}^* R_{\tilde t_i}
        \left[
 L_{\tilde t_i \tilde \chi^0_k}O^{R^*}_{kj}M_{\tilde \chi^+_j} (c_0+c_{11})
-R_{\tilde t_i \tilde \chi^0_k}O^{L^*}_{kj}M_t(c_{12}+c_{23})\right ],
~~~~~~
\end{equation}
where  sums over $i,j,k$ are implied
and the functions $c_{ij}(P_4,P_3,M_{\tilde \chi^+_j},M_{\tilde t_i},
M_{\tilde \chi^0_k})$ are the three-point Feynman integrals[17].
The functions $F^{(ijk)}_{0,1}, G^{(ijk)}_{0,1}$ and constants in the above 
equations are defined as
\begin{eqnarray}
F^{(ijk)}_n&=&\int^1_0 dy y^n\log \left [\frac{m_i^2y(y-1)+m^2_j(1-y)
+m^2_k y}{\mu ^2}\right ],\\
G^{(ijk)}_n&=&-\int^1_0 dy \frac{y^{n+1}(1-y)}{m_i^2y(y-1)+
m^2_j(1-y)+m^2_ky},\\
\lambda_t&=&\frac{M_t}{M_W \sin\beta},\\
L_{\tilde q_1}&=&\cos \theta_{\tilde q},~~
 L_{\tilde q_2}=-\sin \theta_{\tilde q},\\
R_{\tilde q_1}&=&\sin \theta_{\tilde q},~~
 R_{\tilde q_2}=\cos \theta_{\tilde q},\\
L_{\tilde q_1 \tilde \chi^0_j}&=&A_j\cos \theta_{\tilde q}
                        -C_j\sin \theta_{\tilde q},\\
L_{\tilde q_2 \tilde \chi^0_j}&=&-A_j\sin \theta_{\tilde q}
                        -C_j\cos \theta_{\tilde q},\\
R_{\tilde q_1 \tilde \chi^0_j}&=&-A_j^*\sin \theta_{\tilde q}
                        +B_j\cos \theta_{\tilde q},\\
R_{\tilde q_2 \tilde \chi^0_j}&=&-A_j^*\cos \theta_{\tilde q}
                        -B_j\sin \theta_{\tilde q},\\
O^L_{ij}&=&-\frac{1}{\sqrt 2} N_{i4}V^*_{j2}+N_{i2}V^*_{j1},
\end{eqnarray}
and
\begin{equation}
O^R_{ij}=\frac{1}{\sqrt 2} N^*_{i3}U_{j2}+N^*_{i2}U_{j1},
\end{equation}
where $\theta_q$ is the mixing angle of squark $\tilde q$, and
\begin{eqnarray}
A_j&=&\frac{gm_q N^*_{j4}}{2m_W\sin\beta},~~B_j=C_j^*+\frac{gN'_{j2}}{2C_W}\\
C_j&=&\frac{2}{3}eN'^*_{j1}-\frac{2}{3}\frac{gS_W^2}{C_W}N'^*_{j2},\\
N'_{j1}&=&N_{j1}C_W+N_{j2}S_W,\\
N'_{j2}&=&-N_{j1}S_W+N_{j2}C_W.
\end{eqnarray}
In the above $S_W\equiv\sin\theta_W$ and $C_W\equiv\cos\theta_W$.
The chargino masses $\tilde M_j$  and matrix elements $V_{ij}$ depend on
parameters $M,\mu$ and $\tan\beta$, whose expressions can be found in Ref.[11].
The neutralino masses $ M_{\tilde \chi^0_j}$
and matrix elements $N_{ij}$ are obtained by diagonalising the matrix $Y$ [11].
Given the values of the parameters $M, M^{\prime},\mu$ and $\tan\beta$,
the matrix $N$ and $ M_{\tilde \chi^0_j}$ can be obtained numerically. Here,
$\mu$ is the coefficient of
the $H_1-H_2$ mixing term in the superpotential and
$M$ and $ M^{\prime}$ are the masses of gauginos corresponding to $SU(2)$ 
and $U(1)$,
respectively. With the grand unification assumption, i.e. $SU(2)\times U(1)$
is embedded in some grand unified theory, we have the additional relation
$M^{\prime}=\frac{5}{3}\frac{g'^2}{g^2} M$.

The box diagram amplitude $\delta M_{box}$ is 
\begin{eqnarray}
\delta M_{box}&=&i \bar u(P_3)\left [ \left (
  f^b_1 P_R  +f^b_2 P_L +f^b_3 {\large \not} P_4 P_R 
  +f^b_4 {\large \not} P_4 P_L \right ) u(P_1)~
  \bar v(P_2){\large \not} P_3 P_L \right. \nonumber\\
& & \left.+\left ( f^b_5 \gamma^{\mu}P_R 
     +f^b_6 \gamma^{\mu}P_L \right ) u(P_1)~
  \bar v(P_2)\gamma^{\mu}P_L  \right ] v(P_4)
\end{eqnarray}
Here the form factors $f^b_{1,3,5}$ are
\begin{eqnarray}
f^b_1&=&-\frac{g^2}{8\sqrt 2\pi^2}\lambda_t V_{j1}V_{j2}^*
      R_{\tilde t_i}L_{\tilde q_l}L_{\tilde q_l \tilde \chi^0_k}^*
 \left[ L_{\tilde t_i \tilde \chi^0_k}M_t(D_{12}-D_{13}+D_{22}-D_{26})
  \right.       \nonumber\\
& &\left.
 +R_{\tilde t_i \tilde \chi^0_k}M_{\chi^0_k}(D_{12}-D_{13})\right ],\\
f^b_3&=&-\frac{g^2}{8\sqrt 2\pi^2}\lambda_t V_{j1}V_{j2}^*
      R_{\tilde t_i}L_{\tilde q_l}L_{\tilde q_l \tilde \chi^0_k}^*
      L_{\tilde t_i \tilde \chi^0_k}(D_{12}-D_{13}+D_{24}-D_{25}),\\
f^b_5&=&-\frac{g^2}{8\sqrt 2\pi^2}\lambda_t V_{j1}V_{j2}^*
      R_{\tilde t_i}L_{\tilde q_l}L_{\tilde q_l \tilde \chi^0_k}^*
      L_{\tilde t_i \tilde \chi^0_k}D_{27},
\end{eqnarray}
and $f^b_{2,4,6}$ can be obtained through the permutation
\begin{equation}
f^b_{2,4,6}=f^b_{1,3,5}\left \vert_{L_{\tilde t_i \tilde \chi^0_k}
                     \leftrightarrow
                R_{\tilde t_i \tilde \chi^0_k},
        L_{\tilde q_l \tilde \chi^0_k}\rightarrow
                R_{\tilde q_l \tilde \chi^0_k}}.\right.
\end{equation}
The sums over $i,j,k,l$ are implied and
the functions $D_{ij}(P_4,P_3,-P_1,
M_{\tilde \chi^+_j},M_{\tilde t_i},M_{\tilde \chi^0_k},M_{\tilde q_l})$
are the four-point Feynman integrals[17].

The amplitude for the crossed box diagram $\delta M_{box}^c$ is 
\begin{eqnarray}
\delta M_{box}^c&=&-i \bar u(P_3) \left [
  f^c_1 P_L
 +f^c_2 P_R
 +f^c_3 {\large \not} P_4P_L
 +f^c_4 {\large \not} P_4P_R \right ]u(P_2)~
\bar v(P_1) P_L v(P_4),
\end{eqnarray}
where the form factors $f^c_n$ are
\begin{eqnarray}
f^c_1&=&-\frac{g^2}{8\sqrt 2\pi^2}\lambda_t U_{j1}^* V_{j2}^*
      R_{\tilde t_i}L_{\tilde q'_l}L'_{\tilde q'_l \tilde \chi^0_k}
 M_{\chi^+_j}\left[M_t R_{\tilde t_i \tilde \chi^0_k}(D_0+D_{12})
 +M_{\chi^0_k}L_{\tilde t_i \tilde \chi^0_k}D_0\right ],\\
f^c_2&=&-\frac{g^2}{8\sqrt 2\pi^2}\lambda_t U_{j1}^* V_{j2}^*
      R_{\tilde t_i}L_{\tilde q'_l}R'_{\tilde q'_l \tilde \chi^0_k}
 M_{\chi^+_j}\left[M_t L_{\tilde t_i \tilde \chi^0_k}(D_0+D_{12})
 +M_{\chi^0_k}R_{\tilde t_i \tilde \chi^0_k}D_0\right ],\\
f^c_3&=&-\frac{g^2}{8\sqrt 2\pi^2}\lambda_t U_{j1}^* V_{j2}^*
      R_{\tilde t_i}L_{\tilde q'_l}L'_{\tilde q'_l \tilde \chi^0_k}
 M_{\chi^+_j}R_{\tilde t_i \tilde \chi^0_k}(D_0+D_{11}),
\end{eqnarray}
and
\begin{equation}
f^c_4=-\frac{g^2}{8\sqrt 2\pi^2}\lambda_t U_{j1}^* V_{j2}^*
      R_{\tilde t_i}L_{\tilde q'_l}R'_{\tilde q'_l \tilde \chi^0_k}
 M_{\chi^+_j}L_{\tilde t_i \tilde \chi^0_k}(D_0+D_{11}).
~~~~~~~~~~~~~~~~~~~~~~~~~~~~~~
\end{equation}
The sums over $i,j,k,l$ are again implied and
the functions $D_{ij}(P_4,P_3,-P_2,
M_{\tilde \chi^+_j},M_{\tilde t_i},M_{\tilde \chi^0_k},M_{\tilde q'_l})$
are the four-point Feynman integrals[17].
The constants $L'_{\tilde q_i \tilde \chi^0_j},R'_{\tilde q_i \tilde \chi^0_j}$
are defined by
\begin{eqnarray}
L'_{\tilde q_1 \tilde \chi^0_j}&=&A'_j\cos \theta_{\tilde q}
                        -C'_j\sin \theta_{\tilde q},\\
L'_{\tilde q_2 \tilde \chi^0_j}&=&-A'_j\sin \theta_{\tilde q}
                        -C'_j\cos \theta_{\tilde q},\\
R'_{\tilde q_1 \tilde \chi^0_j}&=&-A_j^{\prime *}\sin \theta_{\tilde q}
                        +B'_j\cos \theta_{\tilde q},
\end{eqnarray}
and
\begin{equation}
R'_{\tilde q_2 \tilde \chi^0_j}=-A_j^{\prime *}\cos \theta_{\tilde q}
                        -B'_j\sin \theta_{\tilde q},
\end{equation}
with
\begin{eqnarray}
A'_j&=&\frac{gm_q N^*_{j3}}{2m_W\cos\beta},~~
B'_j=C'^*_j-\frac{gN'_{j2}}{2C_W},
\end{eqnarray}
and
\begin{equation}
C'_j=-\frac{1}{3}eN'^*_{j1}+\frac{1}{3}\frac{gS_W^2}{C_W}N'^*_{j2}.
\end{equation}
\vspace{1cm}

\begin{center} {\Large 3. Numerical results and conclusion }\end{center}

In the following we present numerical results for
the corrections to the total cross section for single
 top quark production via $q \bar q'\rightarrow t \bar b$ at
 the Fermilab Tevatron with $\sqrt s=2$ TeV.
In our numerical calculations we used the MRSG parton distribution
functions[16] and chose the scale $\mu=\sqrt {\hat s}$.
Also we neglected SUSY corrections to the parton distribution functions.
For the parameters involved, we chose $M_Z=91.188 GeV, M_W=80.33 GeV,
M_t=175GeV, M_b=5 GeV$ and $\alpha_{ew}=1/128$.
Other parameters were determined as follows:

(i) The upper bound on $\tan\beta$; viz, $\tan\beta<0.52 GeV^{-1}
 M_{H^+}$, was determined from data on $B\rightarrow \tau \nu X$[18]. 
The lower limits on $\tan\beta$ are $\tan\beta>0.6$ from
perturbative bounds [19] and $\tan\beta>0.25$ (for $M_t=175$GeV) 
from perturbative unitarity[19]. We limited  the value of 
$\tan\beta$ to be in the range of 0.25 to 5, as larger values of 
$\tan\beta$ are not interesting, although allowed by the current data[18],
since the effects are negligibly small.

(ii) For the parameters $M_{\tilde t_R},M_{\tilde t_L},\tan\beta$ and 
$M_{LR}\equiv A_t+\mu\cot\beta$ in top squark (stop) mass matrix[20]
\begin{eqnarray}
M^2_{\tilde t}=\left ( \begin{array}{cc}
M^2_{\tilde t_L}+m_t^2+0.35\cos(2\beta)M_Z^2 &-m_t (A_t+\mu\cot\beta)\\
-m_t (A_t+\mu\cot\beta) & M^2_{\tilde t_R}+m_t^2+0.16\cos(2\beta)M_Z^2
\end{array} \right ),
\end{eqnarray}
 we assumed $M_{\tilde t_R}=M_{\tilde t_L}$. 
There are then three free parameters in the stop sector
and  we chose the mass of the lighter stop $m_{\tilde t_1}, M_{LR}$ and
$\tan\beta$ to be the three independent parameters. 
The best current lower bound on
the stop mass is  55GeV coming from LEP, operating at $\sqrt s=130-140$
GeV[21].
We conservatively took the lower bound to be 50 GeV for  $m_{\tilde t_1}$. 
For the other squarks; i.e., $\tilde q, \tilde q'$ and $\tilde b$,   
we neglected the mixing between left- and right-handed states
and assumed $M_{\tilde q_1}=M_{\tilde q_2}=M_{\tilde q'_1}=M_{\tilde q'_2}=
M_{\tilde b_1}=M_{\tilde b_2}$ which was then determined by[20]
\begin{equation}
m^2_{\tilde b_1}=m^2_b+M^2_{\tilde t_L}
        +\cos(2\beta)(-\frac{1}{2}+\frac{1}{3}S_W^2)M_Z^2
\end{equation}

(iii) For the parameters $M,M', \mu$ and $\tan\beta$ in the 
chargino and neutralino
matrix, we put $M=200$ GeV, $\mu=-100$ GeV and then used the relation
$M'=\frac{5}{3}\frac{g'^2}{g^2} M$ [11] to determine $M'$.

Some typical numerical calculations of the SUSY electroweak corrections are 
given in Figs.2-4.

Fig.2 shows the SUSY electroweak correction $\Delta\sigma/\sigma_0$
as a function of lighter stop mass $M_{\tilde t_1}$,
assuming $\tan \beta=1$ and $M_{LR}=m_t$. The correction is only a few percent
of the tree-level value $\sigma$ and is quite sensitive to the lighter stop 
mass.
There are two peaks at about $M_{\tilde t_1}=75$ GeV and 67GeV due to the fact
that  $m_t=175$ GeV, $M_{\tilde \chi^0_j}=(100, 107, 128, 221)$ GeV
 and the threshold for open top decay into
a neutralino and  a lighter stop is crossed in these regions. 

Fig.3 gives the SUSY electroweak correction 
as a function of $M_{LR}$, assuming $\tan \beta=1$ and $M_{\tilde t_1}=60$ GeV.
This correction is also sensitive to $M_{LR}$. 
With increasing $M_{LR}$ the mass splitting between the 
two stop quarks increases. Since we fixed the mass of $\tilde t_1$,
the mass of $\tilde t_2$ then increases with $M_{LR}$. Furthermore,
with an increase of $M_{LR}$,  $M_{\tilde t_L}$ increases and thus 
the sbottom masses also increase, as seen from Eq.(52).
Since we assumed the masses of the squarks $\tilde q_1, \tilde q_2,\tilde q'_1
$ and $\tilde q'_2$ are degenerate with the sbottoms, these masses then 
also increase  with $M_{LR}$.   
So, with an increase of $M_{LR}$, all the squark masses except the
lighter stop increase and their virtual effects decrease due to decoupling
effects. From Fig.3 one sees that
for $M_{LR}>200$ GeV the magnitude of the correction drops below
one percent.

In Fig.4 we present both the SUSY electroweak correction and the 
Yukawa correction[10] as a function of $\tan \beta$.
For the SUSY electroweak correction we assumed $M_{LR}=m_t$. 
Since both these corrections are proportional 
to $\frac{M_t^2}{M_W^2 \sin^2\beta}$,
they can be very large for small $\tan \beta$.
From Fig.4 one sees that the SUSY electroweak correction exceeds -10\% for
$\tan \beta<0.5$, and tends to infinity when $\tan \beta$ tends to zero.
As in the case of the Yukawa corrections, the SUSY electroweak corrections
are negligibly small for $\tan\beta>1$.
Also, comparing the SUSY electroweak correction with the Yukawa correction
in Fig.4, one notes that the SUSY electroweak
 correction and Yukawa correction have
opposite signs, and thus cancel to some extent.
If the lighter stop quark has the same mass as the charged Higgs boson, the
cancellation is appreciable.
However, as seen in Fig.4, if the charged Higgs is much heavier than the
lighter stop quark, the magnitude of Yukawa correction is much smaller than
the SUSY electroweak correction and there is very little cancellation.
In such a  case the combined effects can exceed -10\% for
$\tan\beta<1$.

To summarize, 
the combined effects of SUSY electroweak
corrections and the Yukawa corrections
can exceed 10\%  for favorable values of the parameters. 
Since the cross section for single top production can be reliably
predicted in the SM [9] and 
the statistical error in the measurement of the cross section  
will be about 6\% at a high-luminosity Tevatron[9], 
these corrections may be observable; at the least, interesting new 
constraints on these models can be established.
\vspace{.5cm}

This work was supported in part by the U.S. Department of Energy, Division
of High Energy Physics, under Grant No. DE-FG02-91-ER4086.
\eject
 
{\LARGE References}
\vspace{0.2cm}
\begin{itemize}
\begin{description}
\item[{\rm [1]}] CDF Collaboration, Phys.Rev.Lett. {\bf 74}, 2626(1995);\\
                 D0 Collaboration, Phys.Rev.lett. {\bf 74}, 2632(1995).
\item[{\rm [2]}] S.Willenbrock, hep-ph/9611240, and reference therein.
\item[{\rm [3]}] F.Berends, J. Tausk and W. Giele, Phys.Rev.D47, 2746(1993).
\item[{\rm [4]}] C.-P. Yuan, hep-ph/9604434, and reference therein.
\item[{\rm [5]}] Dan Amidei and Chip Brock, "Report of the TeV2000 Study
                 Group on Future ElectroWeak Physics at the Tevatron",
                 1995; and reference therein.
\item[{\rm[6]}] S.Dawson, Nucl.Phys. B249, 42(1985)
                S.Willenbrock and D.Dicus, Phys.Rev.D34, 155(1986);\\
                S.Dawson and S.Willenbrock, Nucl.Phys.B284, 449(1987);\\
                C.P.Yuan, Phys.Rev.D41, 42(1990);\\
                F.Anselmo, B.van Eijk and G.Bordes, Phys.Rev.D45, 2312(1992);\\
                R.K.Ellis and S.Parke, Phys.Rev.D46,3785(1992);\\
                D.Carlson and C.P.Yuan, Phys.Lett.B306,386(1993);\\
                G.Bordes and B.van Eijk, Nucl.Phys.B435, 23(1995);\\
                A.Heinson, A.Belyaev and E.Boos, hep-ph/9509274.
\item[{\rm [7]}]  S.Cortese and R.Petronzio, Phys.Lett. B306, 386(1993).
\item[{\rm [8]}]  T.Stelzer and S.Willenbrock, Phys.Lett. B357, 125(1995).
\item[{\rm [9]}]  M.Smith and S.Willenbrock, hep-ph/9604223.
\item[{\rm [10]}] C.S.Li, R.J.Oakes and J.M.Yang, hep-ph/9608460, to appear
                 in Phys.Rev.D
\item[{\rm [11]}] H. E. Haber and C. L. Kane, Phys. Rep. {\bf 117}, 75 (1985);\\
                 J. F. Gunion and H. E. Haber, Nucl. Phys. {\bf B272}, 1 (1986).
\item[{\rm [12]}]  S.Ferrara and E.Remiddi, Phys.Lett.B53(1974)347;\\
                  P.Fayet and S.Ferrara, Phys.Rep.C32(1977)No.5.
\item[{\rm [13]}]   C.S.Li, R.J.Oakes and J.M.Yang, work in progress.
\item[{\rm [14]}] A. Sirlin, Phys. Rev. D22(1980)971;\\
                  W.J. Marciano and A. Sirlin, {\it ibid.} 22, 2695(1980);
                                             31,213(E)(1985);\\
                A. Sirlin and W.J. Marciano, Nucl.Phys.B189(1981)442;\\        
                K.I.Aoki et al., Prog.Theor.Phys.Suppl. 73(1982)1.
\item[{\rm [15]}] H.L. Lai et.al., Phys.Rev.D51, 4763(1995).
\item[{\rm [16]}] A.D. Martin, R.G. Roberts and W.J. Stirling,
		 Phys. Lett. B354, 155(1995).
\item[{\rm [17]}] G. Passarino and M. Veltman, Nucl. Phys. B160(1979)151.
\item[{\rm [18]}] ALEPH Coll. CERN-PPE/94-165.
\item[{\rm [19]}] V.Barger, M.S.Berger and P.Ohmann, Phys.Rev.D47, 1093(1993).
\item[{\rm [20]}] J.Ellis and S.Rudaz, Phys.Lett.{\bf B128}, 248 (1983)\\
        A.Bouquet, J.Kaplan and C.Savoy, Nucl.Phys.{\bf B262}, 299 (1985).
\item[{\rm [21]}] L.Rolandi (ALEPH), H.Dijkstra (DELPHI), D.Strickland (L3),
                 G.Wilson (OPAL), Joint Seminar on the First Results of LEP1.5,
                 CERN, December 12, 1995;\\
                 ALEPH Coll., CERN-PPE/96-10, Jan. 1996;\\
                 L3 Coll., H.Nowak and A.Sopczak, L3 Note 1887, Jan. 1996;\\
                 Opal Coll., S.Asai and S.Komamiya, OPAL Physics Note PN-205,
                 Feb.1996.

\end{description}
\end{itemize}
\eject

\begin{center} {\bf Figure Captions} \end{center}
\vspace{.7cm}

Fig.1 Feynman diagrams for the SUSY electroweak (SUSY EW) corrections: 
(a) tree-level, (b)-(h) one-loop corrections.

Fig.2 The SUSY electroweak (SUSY EW) correction 
$\Delta\sigma/\sigma_0$  as a function of 
$M_{\tilde t_1}$, assuming $\tan \beta=1$ and $M_{LR}=m_t$. 

Fig.3 The SUSY electroweak (SUSY EW) correction 
 as a function of $M_{LR}$, assuming $\tan \beta=1$ and $M_{\tilde t_1}=60$ GeV.

Fig.4 The SUSY electroweak (SUSY EW) correction and the 
Yukawa correction as a function of $\tan \beta$.
$M_{LR}=m_t$ was assumed for the SUSY EW correction.

\end{document}